\documentclass[12pt]{article}
\begin{document}
\begin{center}
{\large{\bf Flat rotational curves without dark matter}}

\vspace*{10mm}

{\bf G.P. Pronko}\footnote{e-mail:pronko@ihep.ru}\\
{\it IHEP, Protvino, Moscow reg., Russia\\
Institute of Nuclear Physics, N.C.S.R "Demokritos" , Athens, Greece}

\begin{abstract}
We consider the singular configurations of gravitating gas
\cite{gravgas} which could be used as a model for disk galaxies. The
simplest steady configuration, which corresponds to rotation of stars
around center gives flat rotational curve, provided the density
distrubution has a tail, decaying as $1/r$. This result based
exclusively on Newtonian gravity and does not involve dark matter.
\end{abstract} 
\end{center}

\section{Introduction}

In the present paper we continue started in \cite{gravgas} study of
the dynamics of the system called gravitating gas. This system is a
kind of continuous media which is constituted by the particles
(stars) interacting with each other only gravitationaly i.e.
according Newton's law. As a mechanical system it could be considered
as a continuous limit of the $N$-particles system described by
Hamiltonian:
\begin{eqnarray}\label{1}
H=\sum_{i=1}^{N}\frac{\vec
p{_i}^2}{2m_i}-\sum_{i\not=j}^{N}\frac{G m_i m_j}{|\vec x_i-\vec
x_j|} , 
\end{eqnarray}
where $\vec x_i, \vec p_i$ are canonical coordinates of particles
(stars)
with masses $m_i$, $G$-gravitational constant.

Speaking about galaxy, where $N\sim 10^{13}$ it seems reasonable to
consider for the system (\ref{1}) the limit $N \rightarrow \infty$
(with assumption $m_i=m$) and substitute the configuration of
particles by continuous distribution of its density and velocity, as
it is done in the theory of gas or fluid.

Certainly, we have to have in mind that the resulting system--
gravitating gas is a very strange gas due to attractive interaction
of its constituents. In the usual gas or fluid, the constituents
(atoms or moleculars) have a repulsive interaction and being put in a
volume it spread in space, filling after some time the whole volume
with
uniform density. As it was proved in \cite{gravgas}, the gravitating
gas, because of attractive interaction may form isolated steady
configuration with asymptotically vanishing density. In particularly,
the galaxy named Hoag's object may be an example of such
gravitational soliton \cite{gravgas}.

Here we are going to consider another kind of steady configurations
of gravitating gas which has a singular distribution of density
\begin{equation}\label{2}
\rho (\vec x)=\delta (z)\rho (r, \phi),
\end{equation}
where $z, r,\phi$ are cylindrical coordinates. This kind of solutions
may be considered as an idealized model of disk galaxies.

Apart from the application on the galactic scale these singular
solutions may be used for the explanation of planetary rings such as
Saturn ring with its surprisingly thin structure formed exclusively
by gravity.

\section{Equations of motion}

In fluid (gas) mechanics there are two different pictures of
description.
The first, usually refereed as Eulerian, uses as the coordinates  the
space dependent fields of velocity and density. The second,
Lagrangian
description, uses the coordinates of the
particles $\vec x(\xi_i,t)$  labeled by the set of the parameters
$\xi_i$ (the numbers of particles in (\ref{1})), which could be
considered as the initial positions
$\vec\xi=\vec x(\xi_i,t=0)$  and time $t$. The useful physical
assumption is  that  the
functions $\vec x(\xi_i,t)$ define a diffeomorphism of  $D \subseteq
R^3$
and the inverse functions $\vec \xi(x_i,t)$ should also exist.
\begin{eqnarray}\label{3}
x_j(\xi_i,t)\Big|_{\vec \xi=\vec\xi(x_i,t)}&=x_j,\nonumber\\
\xi_j(x_i,t)\Big|_{\vec x=\vec x(\xi_i,t)}&=\xi_j.
\end{eqnarray}
The density of the particles in space at time $t$ is
\begin{equation}\label{4}
\rho(\vec x,t)=\int d^3 \xi \rho_0(\xi_i)\delta(\vec x-\vec
x(\xi_i,t)),
\end{equation}
where $\rho_0(\xi)$ is the initial density at time $t=0$.
The velocity field  $\vec v$ as a function of coordinates $\vec x$
and $t$
is:
\begin{equation}\label{5}
\vec v(x_i,t)=\dot{\vec x}(\vec\xi(x_i,t),t),
\end{equation}
where $\vec\xi(x,t)$ is the inverse function (\ref{3}). The velocity
also
could be written in the following form:
\begin{equation}\label{6}
\vec v(x_i,t)=\frac{\int d^3 \xi \rho_0(\xi_i)\dot{\vec
x}(\xi_i,t)\delta(\vec
x-\vec x(\xi_i,t))}{\int d^3 \xi \rho_0(\xi_i)\delta(\vec x-\vec
x(\xi_i,t))},
\end{equation}
or
\begin{equation}\label{7} 
\rho (x_i,t)\vec v(x_i,t)=\int d^3 \xi \rho_0(\xi_i)\dot{\vec
x}(\xi_i,t)\delta(\vec x-\vec x(\xi_i,t)).
\end{equation}
Let us calculate the time derivative of the density using its
definition (\ref{4}) :
\begin{eqnarray}\label{8}
&\dot \rho (x_i,t)=\displaystyle\int d^3 \xi
\rho_0(\xi_i)\frac{\partial}{\partial t}
\delta(\vec x-\vec x(\xi_i,t))\nonumber\\
&=\displaystyle\int d^3 \xi \rho_0(\xi_i)\left(-\dot{\vec
x}(\xi_i,t)\right)\frac{\partial}{\partial \vec x}
\delta(\vec x-\vec x(\xi_i,t))\nonumber\\
&=-\frac{\partial}{\partial \vec x}\int d^3 \xi
\rho_0(\xi_i)\dot{\vec x}(\xi_i,t)\delta(\vec x-\vec
x(\xi_i,t))\nonumber\\
&=-\displaystyle\frac{\partial}{\partial \vec x}\rho (x_i,t)
\vec v (x_i,t)
\end{eqnarray}
In such a way we verify the continuity equation of fluid dynamics:
\begin{equation}\label{9}
\dot \rho (x_i,t)+\vec\partial\Bigl(\rho (x_i,t)\vec v
(x_i,t)\Bigr)=0.
\end{equation}.

Using the coordinates $\vec x(\xi_i,t)$ as a configurational space
variables we can write the Lagrangian for the continuous
generalization of the system, described by (\ref{1})
\begin{equation}\label{10}
L=\int d^3 \xi \rho_0(\xi_i)\frac{m\dot{\vec x}^2
(\xi_i,t)}{2}+\frac{G m^2}{2} \int d^3 \xi d^3 \xi'
\frac{\rho_0(\xi_i)\rho_0(\xi'_i)}{|\vec x(\xi_i,t)-\vec
x(\xi'_i,t)|},
\end{equation}
 The equations of
motion, which follow from the Lagrangian (\ref{10}) have the form:
\begin{equation}\label{11}
m\ddot{\vec x} (\xi_i,t)+G m^2\int d^3 \xi'\rho_{0}(\xi'_i)
\frac{\vec x(\xi_i,t)-\vec x(\xi'_i,t)}{|\vec x(\xi_i,t)-\vec
x(\xi'_i,t)|^3}=0.
\end{equation}
Now we need to translate the equations (\ref{11}) on to the language
of Euler variables. For this let us differentiate both sides of the
equation(\ref{6}) with respect to time:
\begin{eqnarray}\label{12} 
&\displaystyle\frac{\partial}{\partial t}\rho (x_i,t)\vec
v(x_i,t)=\int
d^3 \xi \rho_0(\xi_i)\ddot{\vec x}(\xi_i,t)\delta(\vec x-\vec
x(\xi_i,t))\nonumber\\
&+\int d^3 \xi
\rho_0(\xi_i)\dot{\vec
x}(\xi_i,t)\displaystyle\frac{\partial}{\partial
t}\delta(\vec x-\vec x(\xi_i,t)).
\end{eqnarray}
Substituting $\ddot{\vec x}(\xi_i,t)$ from the equation (\ref{11})
and transforming the second term, as we did in derivation of the
equation (\ref{7})we obtain:
\begin{eqnarray}\label{13} 
&\displaystyle  \frac{\partial}{\partial t}\rho (x_i,t)\vec
v(x_i,t)= -\displaystyle\frac{\partial}{\partial x_k}\Bigl(\rho
(x_i,t)
\vec v (x_i,t) v_k(x_i,t)\Bigr)+\nonumber\\ 
&\displaystyle\int d^3 \xi \rho_0(\xi_i)\Bigl[-G m\int d^3
\xi'\rho_{0}(\xi'_i)
\frac{\vec x(\xi_i,t)-\vec x(\xi'_i,t)}{|\vec x(\xi_i,t)-\vec
x(\xi'_i,t)|^3}\Bigr]\delta(\vec x-\vec
x(\xi_i,t))\nonumber\\
\end{eqnarray}
To transform the last integral we first perform integration over the
$\xi$:
\begin{eqnarray}\label{14}
&\displaystyle\int d^3 \xi \rho_0(\xi_i)\Bigl[-G m\int d^3
\xi'\rho_{0}(\xi'_i)
\frac{\vec x(\xi_i,t)-\vec x(\xi'_i,t)}{|\vec x(\xi_i,t)-\vec
x(\xi'_i,t)|^3}\Bigr]\delta(\vec x-\vec
x(\xi_i,t)) \nonumber \\
&=-\displaystyle G m \rho(x_i,t)\int d^3
\xi'\rho_{0}(\xi'_i)
\frac{\vec x(\xi_i,t)-\vec x(\xi'_i,t)}{|\vec x-\vec
x(\xi'_i,t)|^3}.
\end{eqnarray}
Now let us insert in the integral over $\xi'$ the unity
\begin{equation}\label{15}
1=\int d^3 y \delta({\vec y- \vec x(\xi'_i,t)})
\end{equation}
and change the order of integration. Finally we arrive at
\begin{eqnarray}\label{16}
&\displaystyle\int d^3 \xi \rho_0(\xi_i)\Bigl[-G m\int d^3
\xi'\rho_{0}(\xi'_i)
\frac{\vec x(\xi_i,t)-\vec x(\xi'_i,t)}{|\vec x(\xi_i,t)-\vec
x(\xi'_i,t)|^3}\Bigr]\delta(\vec x-\vec x(\xi'_i,t)= \nonumber \\
&=\displaystyle G m\rho(x_i,t)\frac{\partial}{\partial \vec x} 
\int d^3 y \frac{\rho (y_i,t)}{|\vec x-\vec y|} .
\end{eqnarray}
The equation of motion (\ref{11}) in terms of Euler variables takes
the following form:
\begin{eqnarray}\label{17}
&\displaystyle\frac{\partial}{\partial t}\rho (x_i,t)\vec
v(x_i,t)+\displaystyle\frac{\partial}{\partial x_k}\Bigl(\rho
(x_i,t)\vec v (x_i,t)
v_k(x_i,t)\Bigr)= \nonumber \\
&=\displaystyle G m\rho(x_i,t)\frac{\partial}{\partial
\vec x} \int d^3 y \frac{\rho (y_i,t)}{|\vec x-\vec y|}.
\end{eqnarray}
Using the continuity equation we can rewrite this equation in more
simple form:
\begin{eqnarray}\label{18}
\displaystyle\frac{\partial}{\partial t}\vec
v(x_i,t)+\displaystyle v_k(x_i,t)\frac{\partial}{\partial x_k}\vec v
(x_i,t)= G m\frac{\partial}{\partial
\vec x} \int d^3 y \frac{\rho (y_i,t)}{|\vec x-\vec y|}
\end{eqnarray}
But this is possible only if density does not vanish in some domain.
If for example we want to study configurations with density like
given by (\ref{2}), we must keep the equation of motion in the form
(\ref{17}).

Indeed, let us consider the configuration of our system with
velocity and density of the type
\begin{eqnarray}\label{19}
&\rho (\vec x,t)=\delta (z)\rho (x,y,t)\nonumber\\
&v_i=(v_1(x,y,t),v_2(x,y,t),0).
\end{eqnarray}
Substituting (\ref{19}) into (\ref{17}) we obtain:
\begin{eqnarray}\label{20}
&\displaystyle\frac{\partial}{\partial t}
\delta(z)\rho(x,y,t)(v_1(x,y,t),v_2(x,y,t),0)+\nonumber\\
&\displaystyle\frac{\partial}{\partial x_a}\Bigl(\delta(z)\rho(x,y,t)
v_a(x,y,t)(v_1(x,y,t),v_2(x,y,t),0)\Bigr)= \nonumber \\
&\displaystyle G\ m\ \delta (z)\rho
(x,y,t)(\frac{\partial}{\partial x},\frac{\partial}{\partial y},\frac
{\partial}{\partial z} )\int dx'dy' \frac{\rho
(x',y',t)}{\sqrt{(x-x')^2+(y-y')^2+z^2}},
\end{eqnarray}
where index $a$ takes the values $1,2$.
Apparently, the derivative of integral with  in the r.h.s of
(\ref{20}) with respect to $z$ vanishes after multiplication on 
$\delta(z)$, because the integral is even function of $z$, so the
third
component in both sides of (\ref{20}) disappears and after making use
of the continuity equation we obtain
\begin{eqnarray}\label{21}
&\displaystyle\frac{\partial}{\partial t}v_a (x,y,t)+
v_b(x,y,t)\displaystyle\frac{\partial}{\partial x_b}
v_a(x,y,t)= \nonumber \\
& \displaystyle G m\frac{\partial}{\partial x_a}\int dx'dy'
\frac{\rho
(x',y',t)}{\sqrt{(x-x')^2+(y-y')^2}},
\end{eqnarray}
where indexes $a,b$ take the values $1,2$. This equation has to be
completed with continuity equation, which now has the following form:
\begin{equation}\label{22}
\displaystyle\frac{\partial}{\partial
t}\rho(x,y,t)+\displaystyle\frac{\partial}{\partial
x_a}\Bigl(\rho(x,y,t)v_a(x,y,t)\Bigr)=0
\end{equation}
Note, that as it should be anticipated, the equations (\ref{21}) and
(\ref{22}) took the usual form of 2-dimensional gas dynamics, but the
potential inherited its 3-dimensional form.

\section{Simplest steady solution and its rotational curves. }

Here we are going to consider the time independent configurations of
gravitating gas, therefore all time derivatives in the equations
(\ref{21}) and (\ref{22}) will disappear.
\begin{eqnarray}\label{23}
& v_b(x,y)\displaystyle\frac{\partial}{\partial x_b}
v_a(x,y)=G m\frac{\partial}{\partial x_a}\int dx'dy'
\frac{\rho(x',y')}{\sqrt{(x-x')^2+(y-y')^2}},\nonumber\\
&\displaystyle\frac{\partial}{\partial
x_a}\Bigl(\rho(x,y,t)v_a(x,y,t)\Bigr)=0.
\end{eqnarray}
One of the simplest configuration of gravitating gas describes the
rotation of the particles (stars) around central point. For this
configuration we have:
\begin{eqnarray}\label{24}
v_a =v(r)(\frac{y}{r},-\frac{x}{r}),\nonumber\\
\rho=\rho(r),\qquad r=\sqrt{x^2+y^2}.
\end{eqnarray}
This ansatz satisfies the continuity equation and the main equation
give a relation between velocity $v(r)$ and density $\rho(r)$:
\begin{equation}\label{25}
v^2(r)=-G M r \partial_r \int d r'r'\rho(r')t(r,r'),
\end{equation}
where the kernel $t(r,r')$ is given by:
\begin{eqnarray}\label{26}
&t(r,r')=\displaystyle\int^{2\pi}_{0}d \phi \frac{1}{\sqrt{r^2+r'^2-
2 r r'cos \phi}}=\frac{4}{r+r'}{\bf K}(k),\\ 
& \displaystyle k^2=\frac{4 r r'}{(r+r')^2}\nonumber
\end{eqnarray}
and ${\bf K}(k)$ the complete elliptic integral of the 1-st kind.
So, from this solution follows that any distribution of density is
possible and and we have to consider the stability of solution in
order choose $\rho(r)$.

The equation (\ref{25}) shows that is in case of flat disk
configuration of stars the famous Newton's theorem is not valid. Of
course, this theorem which states that \cite{Gallagher}
\begin{equation}\label{27}
v^2(r)=G m\frac{M(r)}{r}
\end{equation}
where $M(r)$ is the total mass inside the orbit, 
is true only for circular orbits in spherically symmetric density
distribution. In our case we can rewrite (\ref{25}) in the following
form using modular transformation of modulus $k$:
\begin{equation}\label{28}
v^2(r)=-G m r \partial_r 
\Bigl(\int^{r}_{0}
d r'r'\rho(r')\frac{4}{r}{\bf K}(\frac{r'}{r})+\int^{\infty}_{r}
d r'r'\rho(r')\frac{4}{r'}{\bf K}(\frac{r}{r'})\Bigr).
\end{equation}
The first term in the r.h.s. has the factor $\frac{1}{r}$, but
dependence of $r$ is present also in the integrands of both terms
(the dependence from the integration limits is cancelled of course).

So, what we can say about typical rotational curves i.e. the
functions $v(r)$ for different densities? Apparently, if the density
is strictly localized within the circle with radius $R$, when, for
$r\gg R$ we obtain the usual result: 
\begin{equation}\label{29}
v^2(r)=G m\frac{M}{r},
\end{equation}
where $M$ is the total mass (which all is in the circle $r=R$).

In the disk galaxies almost all mass is concentrated near the center
decaying exponentially $e^{-r/h}$ with  $h \sim 1-5\quad kpc$
\cite{Gallagher}. In this case (\ref{29}) could be applied for $r \gg
h$. But, in the same time the disk of the galaxy, as it seen is much
bigger then $h$, therefore we may suggest that there is a tail in the
density distribution which drops slowly, e.g. as a power of $r$.

In order to explore the influence of such kind of densities, we need
to know properties of the  kernel $t(r,r')$ in the
integral(\ref{25}). Apparently it is a symmetric function of  $r,r'$
and has the logarithmic singularity when
$r \rightarrow r'$ (the singularity of elliptic integral ${\bf K}(k)$
for $k \rightarrow 1$). Unfortunately there is only one case
when we can express the integral (\ref{25}) in terms of elementary
functions: for density 
\begin{equation}\label{30}
\rho(r)=\frac{c}{r}
\end{equation}
the integral is given by
\begin{equation}\label{31}
 \int d r'r'\rho(r')t(r,r')=-c\ 2\pi\ln\ r+C',
\end{equation}
where $C'$ is a divergent constant, which is inessential for
calculation of $v(r)$:
\begin{equation}\label{32}
v^2 (r)=G\ m\ c\ 2\pi
\end{equation}
So, the rotational curve for $\rho(r)$, produced by the density
(\ref{30}) is flat! Certainly, in order to have flat rotational
curve, it is sufficient to have the behaviour of density given by
(\ref{30}) only asymptotically, starting e.g. from the distances
$\sim 6-8 kpc$. For smaller distance the exponential distribution
prevails, not changing the behavior at large distance.

As it follows from the analysis of the integral in (\ref{25}), the
density (\ref{30}) is the limiting case for the convergence of
integral which defines $v^2(r)$. If asymptotically 
\begin{equation}\label{33}
\rho(r)=\frac{1}{r^{1+\alpha}},\qquad\alpha>0
\end{equation}
then we have decreasing rotational curve.

\section{Discussion. }

The result we obtained in the previous section is not very unexpected
and is the direct consequence of the flat density distribution. In
order to explain it let us derive it in a more traditional way
without referring to gravitational gas, considering the motion of
particle in axially symmetric density distribution $\rho(x_i)=\rho(r,
\cos \theta)$, where $r,\theta,\phi$ now are spherical coordinates.
Then the equation of motion is given by
\begin{equation}\label{34}
\ddot {\vec x} (t)=G m\vec \partial \int dr' r'^2 d\phi'
d\cos\theta'
\frac{\rho(r', \cos \theta')}{\sqrt{r^2+r'^2-2rr'\cos\Theta}},
\end{equation}
where $\Theta$ is the angle between $\vec x$ and $\vec x'$:
\begin{equation}\label{35}
\cos\Theta=\cos\theta \cos\theta'+\sin\theta
\sin\theta'\cos(\phi-\phi')
\end{equation}
Now let us assume that the test particle moves along circular orbit
with angular velocity $\omega(r)$, therefore
\begin{equation}\label{36}
\ddot {\vec x} (t)=-\omega^2 (r)\vec x(t)=-\frac{v^2(r)}{r^2}\vec
x(t),
\end{equation}
from where we obtain:
\begin{equation}\label{37}
v^2(r)=G m r \partial_r \int dr' r'^2 d\phi'
d\cos\theta'
\frac{\rho(r', \cos \theta')}{\sqrt{r^2+r'^2-2rr'\cos\Theta}},
\end{equation}
Using well-known decomposition of the square root in (\ref{34}) we
can expand the r.h.s. of (\ref{37}) in the following form:
\begin{eqnarray}\label{38}
&v^2(r)=G m r\partial_r\displaystyle \sum^{\infty}_{n=0}\Bigl[\int
^{r}_0 dr'
r'^2 d\phi'
d\cos\theta'\rho(r', \cos
\theta')\frac{r'^{n}}{r^{n+1}}P_n(\cos\Theta)\nonumber\\
&+\displaystyle\int ^{\infty}_r dr' r'^2 d\phi'
d\cos\theta'\rho(r', \cos
\theta')\frac{r^{n}}{r'^{n+1}}P_n(\cos\Theta)\Bigr].
\end{eqnarray} 
If the density  is spherically symmetric, the integration over
angles
$\theta'$ and $\phi'$ leaves only the first term in the sum and we
immediately recover Newton's theorem. In any general case, all terms
of the sum are non-zero and we have complicated behaviour of
rotational curve. If for example the density has the form we had
considered above, i.e. $\rho\sim\rho(r)\delta(r cos(\theta))$, and
the orbit lies in the plane $xy$ we are coming to the equation from
the previous section. Certainly, this derivation can not substitute
the whole dynamics of gravitating gas we had considered, because the
potential we used there is created by the same particles, which move
in the potential.

Now we can estimate the  order of density we need to fit the
observation data of the velocities on the flat part of the rotational
curves. For typical disk galaxy these velocities reach the $150-300
km./sec.$ If we shall take as mass of the particle to be equal to the
mass of our Sun ---$2\cdot 10^{30}$ kg, the coefficient $G M 2\pi$
will be $\sim 8,4\cdot10^{20}m^3/sec^2$. From here  we obtain for the
constant $c$ the values $(0,27-1,1)\cdot10^{-10}m^{-1}$, which
corresponds to $(0,8-3,3)\cdot10^{6}pc^{-1}$. The surface density
therefore becomes
\begin{equation}
\rho(r)=\frac{1}{r}\times(0,8-3,3)\cdot10^{6}pc^{-2},
\end{equation}
where $r$ is measured in pc. On the distance of e.g. 10 kpc, this
formula gives us the surface density $\sim (0,8-3,3)\cdot10^{2}$
stars on square parsec. These numbers is not in contradiction with
observational results \cite{Takamiya}.

The precise measurement of velocities of stars in various galaxies
which was done during last 30 years \cite{Rubin1},\cite{Rubin2} made
it possible to discover the peculiar behaviour of the rotational
curves. This in
turn led one part of community of astronomers and astrophysicists to
the hypothesis of dark matter, the other  to modification of
gravitational law \cite{Bekenstein},\cite{Moffat}. Later another
confirmation of the existance of dark matter has emerged: the data on
gravitational lensing and cosmic microwave background radiation.
However, the argument, given by the  rotational curves stays aside as
a completely model independent measurement, not based on any
assumptions of global structure of Universe. That is the reason why
it is so important to explore all possibility within traditional
physics and I hope that the present paper is one more step in this
direction.

\section{Acknowledgments}

I would like to thank my friends professors A. Likhoded and A.
Rasumov for their interest and support. The work on this paper was
supported in part by the Russian Science Foundation Grant
04-01-00352 and Enter 2004 programme.

\end{document}